\documentclass[aps,groupedaddress]{revtex4}
\usepackage{epsfig}
\usepackage{graphicx}
\usepackage[T1]{fontenc}
\usepackage{ae}
\usepackage{color}
\usepackage[latin1]{inputenc}
\usepackage{amssymb,amsbsy,amsmath}
\usepackage{bbm}
\usepackage{hyperref}

\begin{document}
\bibliographystyle{ieeetran}



\title[Magnetization curves]
{Exotic magnetization curves in classical square-kagom\'{e} spin lattices}

\author{Heinz-J\"{u}rgen Schmidt}
\email{hschmidt@uos.de}
\affiliation{Fachbereich Mathematik/Informatik/Physik, Universit\"{a}t Osnabr\"{u}ck, 49069 Osnabr\"{u}ck, Germany.}

\author{Johannes Richter}
\affiliation{Institut f\"ur Physik, Universit\"at Magdeburg, P.O. Box 4120, D-39016 Magdeburg, Germany}
\affiliation{Max-Planck-Institut f\"{u}r Physik Komplexer Systeme,
        N\"{o}thnitzer Stra{\ss}e 38, D-01187 Dresden, Germany}


\begin{abstract}
Classical spin systems with non-coplanar ground states typically exhibit nonlinear magnetization curves characterized by kinks and jumps. Our article briefly summarizes the most important related analytical results. In a comprehensive case study, we then address AF-square kagom\'{e} and AF/FM-square kagom\'{e} spin lattices equipped with additional cross-plaquette interactions. It is known that these systems have non-coplanar ground states that assume a cuboctahedral structure in the absence of a magnetic field. When a magnetic field $H$ is switched on, a rich variety of different phases develops from the cuboctahedral ground state, which are studied in their dependence on $H$ and a cross-plaquette coupling constant $J_3>0$.
For the AF square-kagom\'{e} spin lattice, we carefully identify and describe seven phases that appear in a phase diagram with five triple points. The transitions between these phases are predominantly discontinuous, although two cases exhibit continuous transitions. In contrast, the phase diagram of the AF/FM square-kagom\'{e}
model shows only four phases with a single triple point,
but these also lead to exotic magnetization curves.
Here, too, there are two types of phase boundaries belonging to continuous and discontinuous transitions.
\end{abstract}

\maketitle

\section{Introduction}\label{sec:I}

Classical Heisenberg spin models with exclusively non-coplanar ground states are attractive for various reasons.
These non-coplanar ground states break a chiral $\mathbb{Z}_2$ symmetry,
leading to phase transitions at finite temperature, even in two space dimensions,
while conventional magnetic order is,
for the Heisenberg models of interest here,
subject to the Mermin-Wagner theorem \cite{mermin1966absence}.
Another motif to study such systems is the possible
route to arriving at quantum spin liquids via frustration-induced quantum melting of magnetic orders,
see \cite{Hickey2017}.

Examples of spin systems with non-coplanar ground states can be found in the recent literature.
A non-coplanar ground state called {\em cuboc2} was found in
Ref.~{\cite{Domenge2005_cuboc2}} for the classical AF kagom\'{e}
with ferromagnetic next-neighbor exchange $J_1$ and antiferromagnetic $2$nd-neighbor exchange $J_2$.
The analogous {\em cuboc1} phase was first reported in Refs.~\cite{janson2008modified,janson2009intrinsic}
for the classical AF kagom\'{e} with antiferromagnetic next-neighbor exchange $J_1$
and $3$rd-neighbor exchange $J_d$ along the diagonals of the hexagons.
Later, these phases were systematically analyzed in
Ref.~\cite{Messio_2011_clas_GS}, and the notations {\em cuboc1} and {\em cuboc2} were introduced
to be explained in more detail in the next paragraph.

A similar spin system on a lattice with corner-sharing triangles is
the square-kagom\'{e} Heisenberg antiferromagnet (SKHAF), that
was introduced about $20$ years ago
\cite{Sidd2001,schnack2001independent,tomczak2003specific,richter2004-spin-peierls,richter2009squago,Sakai2013,
Rousochatzakis2013,derzhko2014square,Rousochatzakis2015,Sakai2015,Derzhko2015review}.
Over the past five years, this system has attracted increasing attention in
theoretical
\cite{Hasegawa2018,Morita2018,Lugan2019,McClarty2020,PhysRevB.102.241115,Iqbal2021,schmoll2022tensor,
squago_j1j2_2023,squago_clas_2023,decorated_squago_2023}
as well as experimental
\cite{FMM:NC20,YSK:IC21,Vasiliev2022,Vasiliev2023,Murtazoev-2023} investigations.

A recent study \cite{squago_clas_2023} focuses on the SKHAF
with longer-range Heisenberg couplings $J_\times$ and $J_+$ across the octagonal plaquettes,
see Fig.~\ref{FIGSQ}.
The spin sites of the square-kagom\'{e} lattice can be divided into two non-equivalent classes,
class A sites, which form the squares, and class B sites,
which are in the middle of the bow ties connecting the squares, see Fig.~\ref{FIGSQ}.
The couplings between the A spins in the squares are denoted by $J_1$  and
the couplings between the A spins and the B spins by $J_2$.
For fully antiferromagnetic (AF) bonds, the spin vectors of the classical ground state
form a {\em cuboctahedron}, one of the $13$ Archimedean solids.
This ground state shows angles of $120^\circ$ between
neighboring spin  and is called a {\em cuboc1 state}.
By inverting the sign of $J_2$
one obtains a variant of the square-kagom\'{e} lattice called AF/FM square-kagom\'{e}.
Its ground state will be the so-called {\em cuboc3 state} obtained by flipping the
B spins of the  cuboc1 state.

In this work, we will continue the investigations of \cite{squago_clas_2023}
by considering the ground states of the AF and AF/FM square-kagom\'{e} lattice
in the presence of a magnetic field $(0,0,H)$ giving rise to an additional Zeeman term
in the Hamiltonian.
Important characteristics to be studied are the magnetization curves
$M(H)$ for different values of the coupling constants,
$S_i^z-$plots of the $z$ components of the ground state spins as a function of $H$,
and the phase diagrams showing different types of ground states
as a function of $H$ and the coupling constants.
To keep the representation simple, we have largely restricted ourselves to the case of
$J_1=1$, $J_2=\pm 1$ and $J_\times =J_+=:J_3>0$.
We then have two phase diagrams corresponding to $J_2=\pm 1$ in the $H - J_3-$ plane.

The reason why we expect interesting magnetization curves for a spin system
with a non-coplanar ground state is as follows.
A classical spin system with a coplanar ground state with vanishing magnetization $M=0$ at $H=0$
always has a linear magnetization curve $M(H)$ with ground states approaching the fully aligned FM state,
similar to the struts of an umbrella when it is folded.
For a spin system with a non-coplanar ground state with $M=0$ for $H=0$,
the analogous construction would also lead to a linear magnetization curve,
but corresponding to a $4-$ dimensional and thus unphysical family of umbrella states.
The necessity of a physical restriction to at most $3-$ dimensional ground states,
on the other hand, leads to a non-linear magnetization curve that typically exhibits jumps and kinks.

In fact, such ``exotic" magnetization curves have so far been described mainly
for spin systems with non-coplanar ground states for $H=0$, see
\cite{Coffey_Trugman_92,SSS:PRL05,Kon:PRB05,Kon:PRB07,Konstantinidis2018,Konstantinidis2023}.
Recently in Ref.~\cite{ZGZ:AP22} unconventional
classical nonlinear magnetization curves have been found for a
frustrated spinel and the $J_1$-$J_d$  AF kagom\'{e}.

The paper is organized as follows. In Section \ref{sec:M} we briefly describe the
numerical, semi-analytical and analytical methods used in the paper.
Section \ref{sec:AF} is devoted to the study of the AF  square-kagom\'{e}.
We find seven phases,
additional to the cuboc1 state for $H=0$ and the FM phase for the saturation regime,
that are described in Section \ref{sec:AF1} together with
typical magnetization curves and an $S_i^z-$plot for $J_3=0.1$.
All phases are non-coplanar like the cuboc1 ground state for $H=0$.
There are discontinuous as well as continuous phase transitions.
Phase IV, which is adjacent to the FM phase,
can be described analytically by a parameter representation,
as explained in the Appendix \ref{sec:AIV}.
For $J_3 \lesssim 0.14$ there is a remarkable phase VII, different from the other ones,
with a large, possibly macroscopic degeneracy.

The phase diagram for the AF/FM  square-kagom\'{e} model is somewhat simpler,
consisting only of four phases, see Section \ref{sec:AFM}.
Also here all phases a non-coplanar, except the analytical phase IV adjacent to the FM regime.
Moreover, the magnetization curve in phase IV is linear, whereas all other
parts of $M(H)$ are nonlinear.

The above-mentioned results are mainly obtained numerically or semi-analytically.
In the section \ref{sec:SR} and in the appendix \ref{sec:SRG} we present some
analytical results for the area under the (normalized) magnetization curve (``sum rule"),
which are valid for general Heisenberg spin systems.
Some of these results are already known, but are scattered in the literature.
For example, this area $F$ is always equal to or larger than the area ${\scriptsize \frac{1}{2}}H_{\rm sat}$
given by the linear magnetization curve.
The latter case, i.e. $F>{\scriptsize \frac{1}{2}}H_{\rm sat}$, which is typical for exotic magnetization curves,
can be predicted by an analysis of the matrix of coupling constants.
Related analytical results on the saturation field
$H_{\rm sat}$ are contained in Appendix \ref{sec:SF}.
We close with a Summary and Outlook in Section \ref{sec:SO}.

\section{Methods}
\label{sec:M}

Let us briefly describe the used methods.
Similar as in Ref.~\cite{squago_clas_2023}
we use a Monte-Carlo like approach to get numerical data for the energy
$E_0$, the magnetization $M$
and spin-spin correlations ${\mathbf s}_i\cdot{\mathbf s}_j$ in the classical ground state.

We also use a semi-analytical approach introduced in \cite{squago_clas_2023}, which works as follows:
Starting from a ground state for a system of $N$ classical spins ${\vec s}_i$,
which we obtain numerically from our Monte Carlo calculation,
we form groups of spin vectors pointing approximately in the same direction, i.~e.,
satisfying, say, ${\mathbf s}_i\cdot{\mathbf s}_j \ge 0.995$.
This gives a set of $M$ distinct spin directions, which is further reduced by guessing their symmetry group.
In the end, we have $K$ different spin directions for the ground state,
of which we obtain all others by applying symmetry operations.
Next, we calculate the energy (including the Zeeman term) of the spin configuration
${\mathcal K}(\alpha_1,\ldots,\alpha_n)$  as a function of some
parameters $\alpha_1,\ldots,\alpha_n$ describing the position of the remaining $K$ spin vectors.
This energy is minimized numerically, starting with the initial
numerical values of the parameters corresponding to the $K$ spin vectors.
It should have become clear that the semi-analytical method cannot be applied schematically,
but requires a certain intuition.
Faulty identifications of closely neighboring spins usually show up in too high ground state energies.
Conversely, a slight lowering of the initial numerical energy is an indication of a successful application of the method.

In special cases, typically in the last phase before saturation, it is possible to solve the system of equations
$0=\partial {\mathcal K}/\partial \alpha_i,\,i=1,\ldots,n$  analytically
and thus obtain explicit or parametric expressions for the magnetization curve.
Alternatively, we used a generalized Luttinger-Tisza method \cite{clas_GS_JPA_2022}
to obtain analytical results for the latter case.
This method provides rigorous results and it can be used to confirm the
semi-analytical results.

\section{AF square-kagom\'{e}}
\label{sec:AF}

\begin{figure}[ht!]
\centering
\includegraphics*[clip,width=0.7\columnwidth]{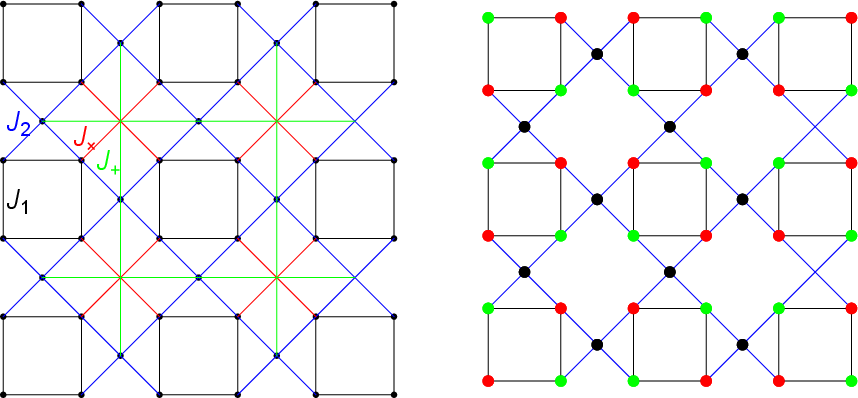}
\caption{Left panel: The square-kagom\'{e} $J_1$-$J_2$ model with cross-plaquette interactions
$J_+$ and $J_{\rm x}$. We distinguish between sites A forming the squares (black lines) and sites B at the center of the
bow ties (blue lines).\\
Right panel: Sketch of phase IV for the AF/FM square-kagom\'{e} lattice, see Section \ref{sec:AFM24}. The black dots at sites B represent spin vectors of the form
${\mathbf s}_i=(0,0,1)$, and the red/green dots at sites A  spin vectors of the form
${\mathbf s}_i=(\pm\sqrt{1-z_1^2},0,z_1)$, where $z_1$ is
given below, see Eq.~(\ref{ph4z}).
}
\label{FIGSQ}
\end{figure}
Like the kagom\'{e} lattice the square-kagom\'{e} consists of corner-sharing triangles,
but enclosing squares instead of hexagons.
Another difference to the kagom\'{e} lattice is
the existence of two non-equivalent sites
A (forming the squares) and B (center of the bow ties)
as well as two non-equivalent nearest neighbor bonds $J_1$ and $J_2$.
The corresponding model is depicted in Fig.~\ref{FIGSQ}, left panel.
For numerical and some analytical calculations we have to approximate the infinite
square-kagom\'{e} lattice by a finite lattice model ${\mathcal L}$ with periodic boundary conditions.
We may view ${\mathcal L}=\left({\mathcal V}, {\mathcal E}\right)$ as an undirected
graph with vertex set ${\mathcal V}=\{1,\ldots,N\}$ representing the spin sites and a set of edges
\begin{equation}\label{edges}
 {\mathcal E} = {\mathcal SQ}\cup {\mathcal BT}\cup {\mathcal C}\cup {\mathcal P}
 \;,
\end{equation}
representing, resp.,  the interacting pairs within the squares, bow ties, the diagonal cross-plaquette pairs, and the vertical/horizontal ones.
Then the Hamiltonian ${\mathcal H}$ (without Zeeman term) can be written as
\begin{equation}\label{Hamiltonian}
{\mathcal H}=\sum_{(i,j)\in {\mathcal SQ} }J_1\,{\mathbf s}_i\cdot{\mathbf s}_j+
\sum_{(i,j)\in {\mathcal BT} }J_2\,{\mathbf s}_i\cdot{\mathbf s}_j+
\sum_{(i,j)\in {\mathcal C} }J_\times\,{\mathbf s}_i\cdot{\mathbf s}_j+
\sum_{(i,j)\in {\mathcal P} }J_+\,{\mathbf s}_i\cdot{\mathbf s}_j
\;,
\end{equation}
see Fig.~\ref{FIGSQ}, left panel.

Recall that the zero-field cuboc ground-state phase is present in the entire
parameter region $J_+,J_{\rm x}>0$ independent of the magnitudes of $J_+$ and $J_{\rm
x}$, see Fig.~5 in Ref.~\cite{squago_clas_2023}.
For the sake of simplicity we therefore consider only the symmetric case
$J_+=J_{\rm x}=J_3>0$ and further set $J_1=J_2=1$.

\subsection{Phase diagram}
\label{sec:AF1}

\begin{figure}[ht!]
\centering
\includegraphics*[clip,width=1.0\columnwidth]{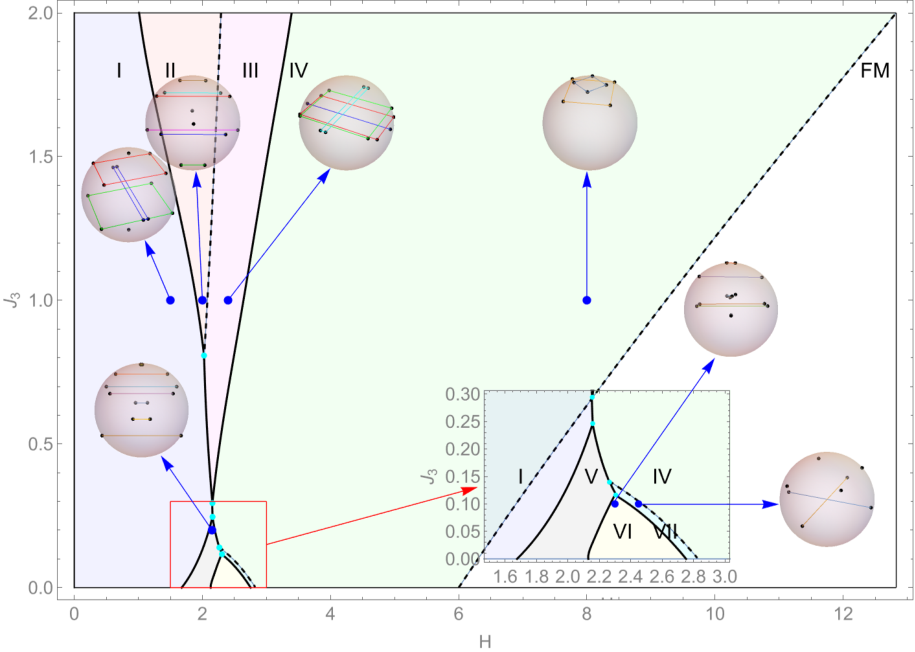}
\caption{Phase diagram for the AF square-kagom\'{e} model in the $H-J_3$-plane,
where $H$ denotes the magnetic field and $J_3$ the additional cross-plaquette coupling.
We have chosen $0\le J_3\le 2$ and $0\le H\le 12.8284...$.
There are seven phases  I - VII, additional to the cuboc1 ground state for $H\to 0$
and the ferromagnetic ground state FM for $H\ge H_{\rm sat}$ and five triple points (cyan color).
The solid curves represent discontinuous phase transitions and the dashed
lines continuous ones.
For the region  $1.5 \le H \le 3$ and $0\le  J_3\le 0.3$ we have added an enlargement in the form of an inset.
Typical common-origin plots for all seven phases have been inserted such that the corresponding point in the
$H-J_3$-plane is indicated by a blue dot.
}
\label{FIGPD1}
\end{figure}

\begin{figure}[h!]
\centering
\includegraphics*[clip,width=0.8\columnwidth]{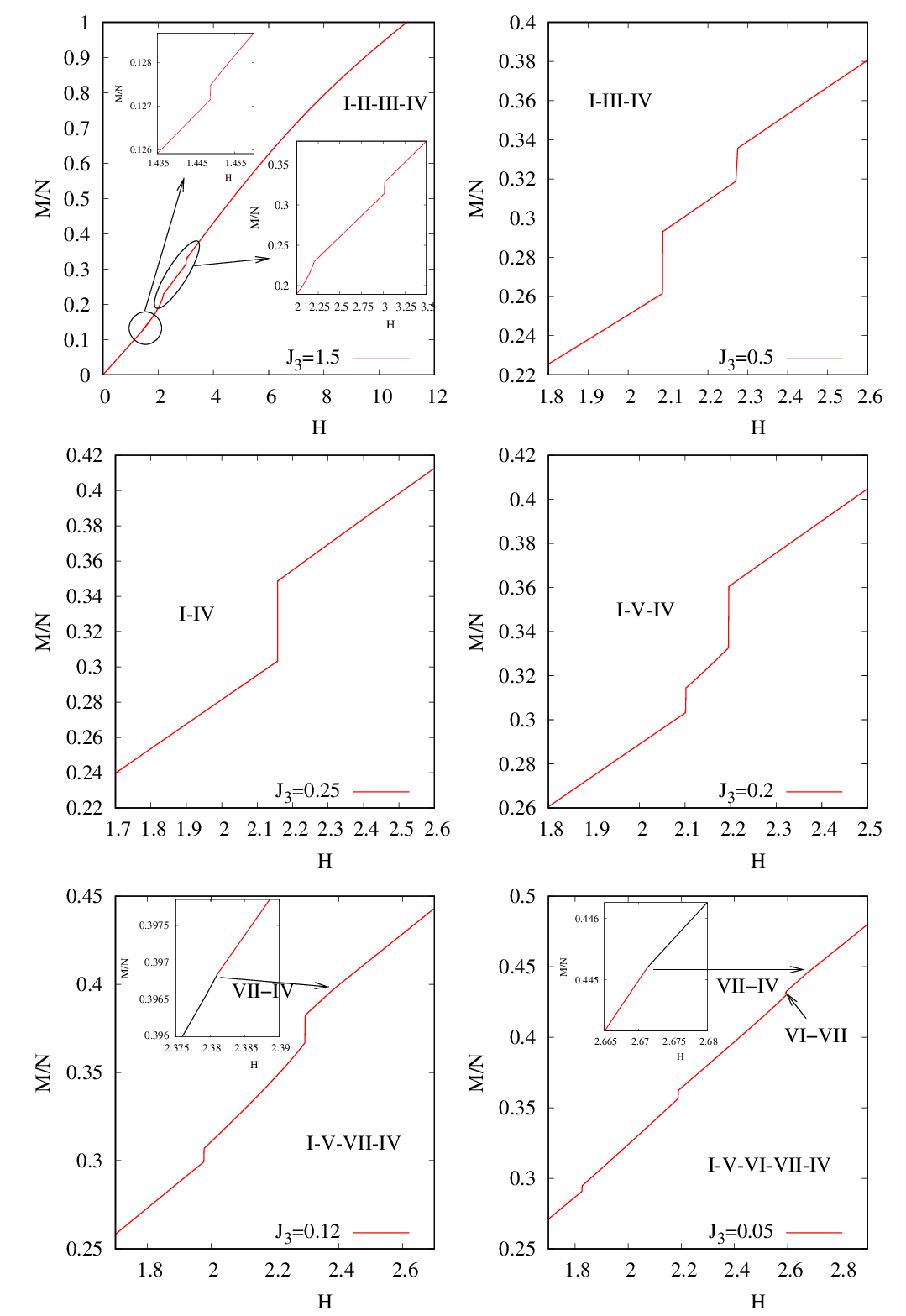}
\caption{According to Table \ref{tab1} and the phase diagram in Figure \ref{FIGPD1}
there are six different types of magnetization curves,
of which we show typical sections that were calculated numerically.
From top to bottom, $J_3$ assumes the values $J_3=1.5, 0.5, 0.25, 0.2, 0.12, 0.05$.
In the cases where the magnetization jumps or kinks are not visible,
we have inserted enlargements in the form of insets.
The two insets for $J_3=0.12$ and $J_3=0.05$ are based on semi-analytical calculations.
}
\label{FIGmagafm}
\end{figure}

In the phase diagram (Fig.~\ref{FIGPD1}) we have chosen $0\le J_3\le 2$ and $0\le H \le 10 + 2 \sqrt{2} =12.8284...$
according to the value of $H_{\rm sat}=4 + 3 J_3 + \sqrt{4 + J_3^2}$, see (\ref{hsatSK3}),
for $J_3=2$. There are seven phases I - VII, additionally to the cuboc1 ground state for $H\to 0$
and the ferromagnetic ground state FM for $H\ge H_{\rm sat}$.
We obtain two continuous transitions (additional to the transition IV-FM),
corresponding to the phase boundary between IV and VII and the phase boundary between II and III.
In both cases of continuous transitions, we have a subgroup relationship 
between the symmetry groups of the phases involved, as it must necessarily be.
The magnetization jump $\Delta M$ at the phase boundary between I and II varies over three
decades and, for $J_3\gtrsim 1.6$, is hardly visible in the numerically calculated magnetization curves.
We observe five triple points $T_i, i=1,\ldots,5$, where three phase boundaries
intersect and three phases touch, see the cyan points in Fig.~\ref{FIGPD1} and Table \ref{tab1}.
As a consequence there are six types of magnetization curves, see Table \ref{tab1} and Fig.~\ref{FIGmagafm}.

\begin{table}
  \centering
  \caption{Coordinates of the five triple points $T_i,\, i=1,\ldots,5$ in the phase diagram (Fig.~\ref{FIGPD1})
  and the type of magnetization curves for values of $J_3$ between the triple points.
  When listing the jumps and kinks in the magnetization curve
  we always neglect the kink at $H=H_{\rm sat}$. }\label{tab1}
  \begin{tabular}{|c|c|c|c|l|c|}
    \hline
    $i$ & $H=h_i$ & $J_3=j_i$ && interval & jumps (${\mathcal J}$) and kinks (${\mathcal K}$) \\
    \hline
    1 & 2.02593 & 0.80768 &&$j_1<J_3$ & ${\mathcal J}\,{\mathcal K}\,{\mathcal J}$\\
    2 & 2.1556 & 0.294157 &&$j_2<J_3<j_1$ & ${\mathcal J}\,{\mathcal J}$\\
    3 & 2.15822 & 0.246352 &&$j_3<J_3<j_2$ & ${\mathcal J}$ \\
    4 & 2.26488 & 0.13962 &&$j_4<J_3<j_3$ & ${\mathcal J}$\, ${\mathcal J}$\\
    5 & 2.30494 & 0.116264 &&$j_5<J_3<j_4$ & ${\mathcal J}$\, ${\mathcal J}$\, ${\mathcal K}$\\
    &&&& $ 0<J_3<j_5 $& ${\mathcal J}$\, ${\mathcal J}$\, ${\mathcal J}$\, ${\mathcal K}$\\
   \hline
  \end{tabular}
\end{table}

\subsection{Description of single phases}
\label{sec:AF2}

Here we describe the seven phases of the phase diagram in Fig.~\ref{FIGPD1}.
The corresponding ``common-origin plots" in this Figure provide a graphical
illustration of their spin structure.
In a common-origin plot, the entire ground state configuration of spins in real space
is concentrated into a point cloud on a single unit sphere.
We also show one representative example of the field
dependence of the $z$-components of the spin vectors of the ground state configuration
in Fig.~\ref{FIGSiz}.

\subsubsection{Phase I}
\label{sec:AF21}

We observe $14$ different spin vectors, two of which occupy the north and
south pole of the unit sphere and the remaining $12$ are distributed on three
different circles with constant $z$ in groups of four. These groups form rectangles
with parallel edges. Hence the symmetry group of this phase is $D_2$, the
symmetry group of the rectangle.
For $H\to 0$ the points of the center rectangle merge in pairs, and the resulting
$12$ spin vectors form a cuboctahedron.
We show an example corresponding to $ H=1.5$ and $J_3=1$, see Fig.~\ref{FIGPD1}.
In every finite model of the square-kagom\'{e} lattice with $N$ spin sites,
$N=6\,L^2,\; L$ even,
$10$ spins occupy sub-lattices with $N/12$ sites and four spins occupy
sub-lattices with $N/24$ sites.
The partial merging for $H\to 0$ results in the $12$
sublattices of the cuboc1 state, each with $N/12$ sites.

\subsubsection{Phase II}
\label{sec:AF22}

We observe $14$ different spin vectors, which are pairwise distributed on six
different circles with constant $z$ and two further spin vectors with different $z$,
see the example in Figure \ref{FIGPD1} for $H=1.5$ and $J_3=1$.
There is a reflection symmetry which can be described as $y \leftrightarrow-y$ 
upon a suitable choice of the coordinate system and hence the symmetry group is $D_1$.

\subsubsection{Phase III}
\label{sec:AF23}

If $H$ is increased beyond a critical value, the eight circles of phase II
merge into four circles with constant $z_i,\,i=1,\ldots,4,$ and thus form a continuous transition to another phase called III.
Three of these circles contain rectangles with parallel edges formed by spin vectors and
the fourth circle contains a pair of antipodal spin vectors, also parallel to the edges of
the rectangles, see the example in Figure \ref{FIGPD1} for $H=2.4$ and $J_3=1$.
This spin configuration has thus $D_2$-symmetry.
Since $z_1$ and $z_3$ are very close to each
other, the corresponding circles cannot be distinguished by eye.

\subsubsection{Phase IV}
\label{sec:AF24}

Further  increasing $H$ (e.~g.~beyond $H=2.8$ at $J_3=1.0$)  we obtain a new phase denoted by IV.
It consists of eight different spin vectors, which are distributed on two circles with
constant $z=z_i,\,i=1,2$,  see the example in Figure \ref{FIGPD1} for $H=8$ and $J_3=1$.
The two groups of spin vectors form two squares twisted by $45^\circ$.
Hence this ground state has a $D_4$ symmetry.

This phase can be analytically described by a parameter
representation. Relevant quantities as magnetic field, $z_1$, $z_2$,  magnetization and energy
can be expressed as functions of a parameter $w$ that varies between $1$ and $w_{max}(J_3)$.
These functions are specified in Appendix \ref{sec:AIV}.

\subsubsection{Phase V}
\label{sec:AF25}

For small values of $J_3$, by increasing $H$ we enter a new phase  V directly
from phase I (e.~g.~for $J_3=0.2$ at $H = 2.10188$), where phase V is similar to phase
II.
We observe $14$ different spin vectors, which are pairwise distributed on seven
different circles with constant $z$,  see the example in Figure \ref{FIGPD1} for $H=2.15$ and $J_3=0.2$.
There is a $D_1$ reflection symmetry which can be described as $y \leftrightarrow-y$ 
upon a suitable choice of the coordinate system.

\subsubsection{Phase VI}
\label{sec:AF26}

Below the triple point $T_5$ phase V gives way for
a new phase IV
(e.~g.~for $J_3=0.1$ at $H=2.387$).
We observe $14$ different spin vectors, which are pairwise distributed on seven
different circles with constant $z$ forming pairs,
see the example in Figure \ref{FIGPD1} for $H=2.3$ and $J_3=0.1$.
The symmetry of this phase is the rotation of the $x-y$-plane with $180^\circ$, or
$(x,y)\leftrightarrow(-x,-y)$ and hence its symmetry group of order $2$ can be chosen as $C_2$
in order to distinguish it from the reflection symmetry group $D_1$ of phase II.

\newpage

\newpage

\subsubsection{Phase VII}
\label{sec:AF27}

\begin{figure}[ht!]
\centering
\includegraphics*[clip,width=1.0\columnwidth]{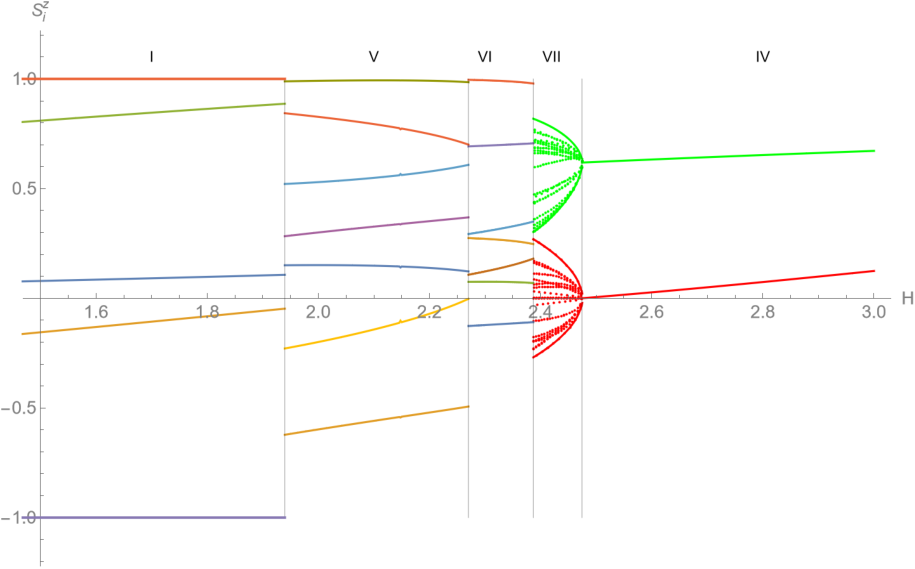}
\caption{The $z$-components of the spin vectors of the ground state configuration semi-analytically calculated for
$J_1=J_2=1$,  $J_3=0.1$
and  $1.5\le H \le 3.0$. We can distinguish all five phases I,V,VI,VII,IV, similarly as in the magnetization
curve, see Figure \ref{FIGmagafm}, lower right panel.
Note the large degeneracy of the ground states of phase VII for $2.387<H<2.47456$, of which we have only drawn a selection.
}
\label{FIGSiz}\end{figure}

For $J_3=0.1$ and $2.387<H<2.47456$ we obtain a new, disordered phase denoted by VII.
It has a large degeneracy, as shown by Figure \ref{FIGSiz}, reminiscent of
the spin liquid for $J_3=H=0$, see \cite{squago_clas_2023}.
For the semi-analytical calculation we have randomly picked out
a variety of ground state configurations of this phase and found eight
different spin vectors, such that four of them are arbitrarily positioned on
the unit sphere and the remaining four spin vectors forming antipodal
pairs, see the example in Figure \ref{FIGPD1} for $H=2.45
$ and $J_3=0.1$. The symmetry group of this phase is thus reduced to the identity.
Both groups of spin vectors have strictly separated $z$-values.
As $H$ increases and approaches the phase boundary between phase VII and IV,
the $z$ values of the first group (with uncorrelated $z$-values) converge towards some value $z_I>0$,
while the $z$ values of the second group (forming antipodal pairs) converge towards $z_{II}=0$,
see Figure \ref{FIGSiz}.
This numerical finding suggests the approach of analytically determining the phase boundary VII/IV
by the condition that one of the two $z$-values in phase IV is equal to $0$. The result

\begin{equation}\label{phaseboundary74}
  J_3=1-\frac{2}{\sqrt{8-\frac{H^2}{2}}}
\end{equation}
agrees very well with the semi-analytically calculated phase boundary.
It follows that the first, non-zero limit value for $z$ is
\begin{equation}\label{zII}
 z_{I}=\frac{\sqrt{\left(J_3-2\right) J_3+\frac{1}{2}}}{1-J_3}
 \;.
\end{equation}
For $J_3=0.1$ used in Fig.~\ref{FIGSiz} we have $z_{I}=0.61864$.
The corresponding phase transition is continuous.

\section{AF/FM square-kagom\'{e}}
\label{sec:AFM}
The AF/FM square-kagom\'{e} is obtained for $J_1>0$ and $J_2<0$, i.~e. for AF couplings within the squares and
FM couplings within the bow ties connecting the squares.
If the magnetic field vanishes, the AF/FM case can be reduced to the
 AF case by simultaneously inverting the sign of $J_2$ and flipping
the spin vectors at the sites B (sitting in the center of the bow ties), leaving the Hamiltonian unchanged.
Consequently, the ground state in the quadrant $J_\times,J_+>0$ will be the cuboc3 spin configuration obtained from the cuboc1 ground state of the AF
square-kagom\'{e} by flipping the spins at sites B, see \cite{squago_clas_2023}.
However, the Zeeman term is not invariant under this transformation, so that the magnetization of the
AF/FM square kagom\'{e} must be investigated separately. For this we set $J_1=1$ and $J_2=-1$ and restrict ourselves to the diagonal $J_\times=J_+=:J_3>0$.

\subsection{Phase diagram}
\label{sec:AFM1}

\begin{figure}[ht!]
\centering
\includegraphics*[clip,width=1.0\columnwidth]{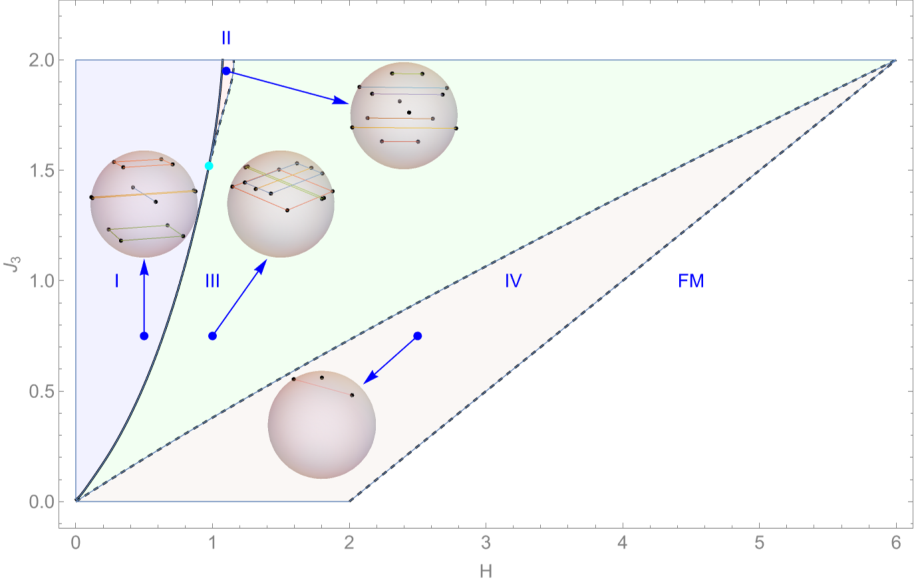}
\caption{Phase diagram for the AF/FM square-kagom\'{e} model in the $H-J_3$-plane,
where $H$ denotes the magnetic field and $J_3$ the additional cross-plaquette coupling.
We have chosen $0\le J_3\le 2$ and $0\le H\le 6$.
In addition to  the cuboc3 ground state for $H\to 0$
and the ferromagnetic ground state FM for $H\ge H_{\rm sat}$ there are four phases  I - IV
and one triple point (cyan color).
The solid curves represent discontinuous phase transitions and the dashed ones continuous phase transitions.
}
\label{FIGPD3}\end{figure}

\begin{figure}[ht!]
\centering
\includegraphics*[clip,width=0.7\columnwidth]{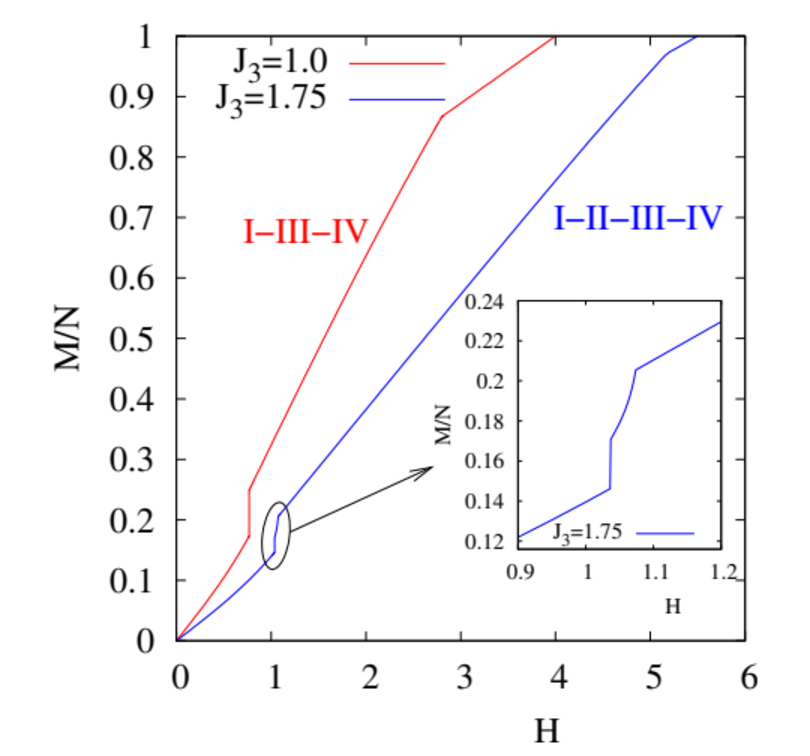}
\caption{According to the phase diagram in Figure \ref{FIGPD3}, there are two different types of magnetization
curves.
We show typical $M(H)$ curves for the values $J_3=1.0$ and $J_3=1.75$ that were calculated numerically.
In the case $J_3=1.75$ we have added an enlargement in the form of an inset to make the jump and the kink
in the magnetization curve more clearly visible.
}
\label{FIGmagfm}\end{figure}

\begin{figure}[ht!]
\centering
\includegraphics*[clip,width=1.0\columnwidth]{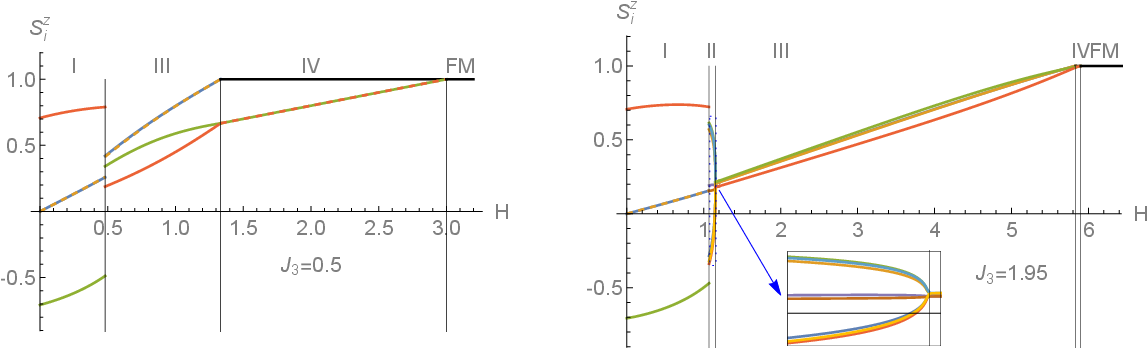}
\caption{The $z$-components of the spin vectors of the ground state configuration semi-analytically calculated for
$J_1=-J_2=1$,  $0\le H \le H_{\rm sat}$. and two different values of $J_3$.\\
Left panel: $J_3=0.5$.
We can distinguish the phases I-III-IV, similarly as in the left magnetization
curve of Fig.~\ref{FIGmagfm}.\\
Right panel: $J_3=1.95$. Here also the phase II is visible and the continuous transition to phase III, see the enlargement in the inset.
}
\label{FIGSIZ2}\end{figure}

In the phase diagram, see Fig.~\ref{FIGPD3}, we have chosen $0\le J_3\le 2$ and $0\le H \le 6$
according to the value of $H_{\rm sat}=2(1+J_3)$, see (\ref{hsatSKF1}),
for $J_3=2$. There are four phases I - IV, additionally to the cuboc3 ground state for $H\to 0$
and the ferromagnetic ground state FM for $H\ge H_{\rm sat}$.
We found two continuous transitions (additional to the transition IV - FM)
corresponding to the phase boundaries between  II and III and between III and IV.
In both cases there is a symmetry group inclusion of the form $D_1 \subset D_2$.
We obtain one triple point at $H=h_1=0.974276, \, J_3=j_1=1.52014$, where the three phases I,II, III touch,
see the cyan point in Fig.~\ref{FIGPD3}.
As a consequence there are two types of magnetization curves,
one with one jump and one kink for $J_3>j_1$ and one with one jump and two kinks for $J_3<j_1$, see  Fig.~\ref{FIGmagfm}.
Two typical plots of the $z$-components of all ground state spin vectors depending on $H$ are shown in Fig.~\ref{FIGSIZ2}.

\subsection{Description of single phases}
\label{sec:AFM2}

Here we describe the four phases of the phase diagram in Figure \ref{FIGPD3}.
The corresponding common-origin plots in this Figure provide a graphical
illustration of their spin structure.

\subsubsection{Phase I}
\label{sec:AFM21}

For small $H$ we observe a ground state configuration, called phase I,
consisting of $14$ spin vectors with four different $z$ values $z_i,\,i=1,\ldots,4$.
They form three rectangles with parallel edges
that can be chosen parallel to the $x$- and $y$-axis
and a pair of the form $(\pm \sqrt{1-z_4^2},0,z_4)$.
The symmetry group of phase I is hence $D_2$.
For $H\to 0$, one of the rectangles degenerates into a pair of spin vectors
and the resulting $12$ vectors form the cuboc3 ground state mentioned above.

\subsubsection{Phase III}
\label{sec:AFM23}

This phase is adjacent to phase I below the triple point,
i.~e., when $J_3<j_1=1.52014$ and has the same structure as phase I, i.~e., 
consisting of three rectangles and one pair, also with the symmetry group $D_2$,
but a different energy.
When approaching the boundary to phase IV, see below, one of the rectangles
and the pair of spin vectors converge towards the north pole $(0,0,1)$
of the unit sphere and the other two rectangles approach two vectors of the form
$(\pm\sqrt{1-z_1^2},0,z_1)$, resulting in a continuous transition to phase IV.

\subsubsection{Phase II}
\label{sec:AFM22}

Above the triple point, i.~e., when $J_3>j_1=1.52014$, there exists another phase, called phase II, located between
phase I and phase III. It consists of $14$ spin vectors forming six pairs related by the symmetry $y \leftrightarrow -y$,
if we choose the coordinated system correspondingly, and two further spin vectors with $y=0$.
The symmetry group is thus $D_1$.

\subsubsection{Phase IV}
\label{sec:AFM24}
This phase is the only one with coplanar spin structure and consequently the
corresponding part of the magnetization curve is linear.
It can be analytically described
and consists of three different sublattices of spin vectors of the form
${\mathbf s}_i=(0,0,1)$ and ${\mathbf s}_i=(\pm\sqrt{1-z_1^2},0,z_1)$
distributed on the spin lattice according to Fig.~\ref{FIGSQ}, right panel. 
Hence phase IV has thus the same symmetry group $D_1$ as phase II.
The various quantities
as $z_1$, magnetization $M/N$ and energy $E$ are given as functions of $H$ and $J_3$ by
\begin{eqnarray}
\label{ph4z}
  z_1 &=& \frac{H+2}{2 \left(J_3+2\right)}\;, \\
  \label{ph4M}
  M/N&=& \frac{H+J_3+4}{3 \left(J_3+2\right)}\;, \\
  \label{ph4E}
  E &=& \frac{(H+2) \left(H+2 J_3+6\right)}{6 \left(J_3+2\right)}
  \;.
\end{eqnarray}
For $H\to H_{\rm sat}=2(1+J_3)$  it follows that $z_1\to 1$ and $M/N\to 1$, i.~e., the ground state approaches the FM state.

\section{Sum rule for magnetization curves}\label{sec:SR}

In this section we will formulate some results on the area below the magnetization curves for
the square-kagom\'{e} systems that we consider in this article and which are of relevance for the exotic
character of these curves.
More general results that hold for general classical Heisenberg spin systems are proven in Appendix \ref{sec:SRG}.

\begin{figure}[ht!]
\centering
\includegraphics*[clip,width=0.9\columnwidth]{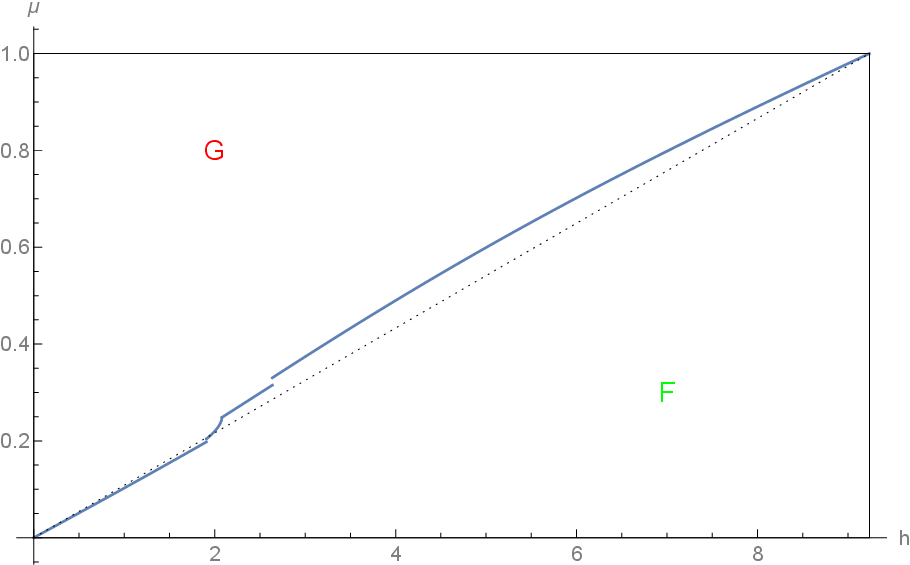}
\caption{Magnetization curve for the AF square-kagom\'{e} and coupling constants $J_1=J_2=J_3=1$,
the area $F$ below the magnetization curve and the complementary area $G$. The area $F\approx 4.90273$
is larger than the area $\frac{1}{2}H_{\rm sat}\approx 4.618$ below the linear magnetization curve (dotted line).
}
\label{FIGMC}\end{figure}

For a linear magnetization curve $M/N= \frac{H}{H_{\rm sat}}$ (dotted line in Figure \ref{FIGMC}) the area $F$ below the
graph of $M/N$ is $\frac{1}{2}H_{\rm sat}$. In general, one can show that $F\ge \frac{1}{2}H_{\rm sat}$, see Appendix \ref{sec:SRG}.
Hence $F> \frac{1}{2}H_{\rm sat}$ implies that the
magnetization curve cannot be completely linear and may show exotic
features as jumps or kinks.

For the AF square-kagom\'{e} with $J_1=J_2=1$ and $J_+=J_\times=:J_3$ we have the following result:
\begin{equation}\label{sumrule1}
  \Delta F := F-\frac{1}{2}\,H_{\rm sat}=
 -1+ \frac{1}{2} \sqrt{J_3^2+4}+\frac{J_3}{6} > 0
   \;,
\end{equation}
for all $J_3>0$ thus indicating exotic magnetization curves.
To give a numerical example: For $J_3=1$, corresponding to Figure \ref{FIGMC}, we have $H_{\rm sat}=12.8284$ and the
numerical integration of the magnetization curve yields $\Delta F=0.2847003 $, very close to the analytical value
of $\Delta F= \frac{1}{6} \left(3 \sqrt{5}-5\right)=0.2847006554\ldots$.

For AF/FM square-kagom\'{e} systems with $J_1=-J_2=1$ and $0\le J_+=J_\times=:J_3 < 2$ the analogous result reads:
\begin{equation}\label{sumrule2}
  \Delta F := F-\frac{1}{2}\,H_{\rm sat}=
  \frac{1}{2}\left( 2-J_3\right) > 0
   \;,
\end{equation}
To give an example: For $J_3=1/2$  we have $H_{\rm sat}=3$
according to $H_{\rm sat}=2\left( 1+J_3\right)$, see (\ref{hsatSKF1}).
The numerical integration of the magnetization curve yields $\Delta F=0.50002 $, very close to the analytical value
of $\Delta F= 1/2$.

\section{Summary and Outlook}\label{sec:SO}
Spin systems with non-coplanar ground states are expected to have non-linear magnetization curves with kinks and jumps.
In general, magnetization curves can only be studied numerically or semi-analytically,
but some features can be determined analytically, e.g. the saturation field,
the area under the magnetization curve and sometimes
the last part of the curve just before saturation.
We have compiled the corresponding analytical results in this paper.

As an extended case study, we have further investigated the
AF square-kagom\'{e} and AF/FM square-kagom\'{e} lattices
which are known to have exclusive non-coplanar ground states with
cuboctahedral structure when the magnetic field vanishes.
Indeed we found a diverse spectrum of ground state phases
depending on the magnetic field $H$ and a certain coupling constant $J_3>0$.
The latter describes the strength of the additional cross-plaquette interactions.

For the AF square-kagome, we were able to identify seven phases with five triple points
and describe their structure in detail.
The transitions between these phases are mostly discontinuous,
but in two cases also continuous.
They can be characterized more precisely with so-called $S_i^z-$plots showing the $z$-components of all ground
state spin vectors as a function of $H$.
In contrast, the AF/FM square kagome has only four phases with one triple point,
but also shows exotic magnetization curves with continuous as well as discontinuous phase transitions.

The question can be raised whether our results are typical for spin systems with cuboc ground states.
As mentioned in the Introduction, cuboc phases were originally found in variants of the AF kagom\'{e} lattice.
Therefore, the investigation of magnetization curves etc.~for these models
would be an obvious extension of the present work. Corresponding studies are in preparation.

\section*{Acknowledgment}\label{sec:ACK}
We thank Martin Gemb\'{e}, Ciar\'{a}n Hickey, Yasir Iqbal and Simon Trebst for instructive discussions
in connection with a common project on the square-kagom\'{e} spin lattices,
from which the topic of the present paper developed.

\appendix

\section{Saturation fields}\label{sec:SF}

When the magnetic field $H$ reaches or exceeds a certain saturation value $H_{\rm sat}$,
all spins align parallel to the direction of the magnetic field.
In calculating $H_{\rm sat}$ for the systems considered in this work
we have followed the approach presented in
\cite{HJS_classical_I,HJS_classical_IV,clas_GS_JPA_2022},
which will be briefly recapitulated here.
This approach is limited to finite spin systems but it turns out that the results for $H_{\rm sat}$ are independent of the
system size (provided $N$ is not too small) and hence also hold for the infinite square-kagom\'{e} spin lattice.

Thus we consider the matrix ${\mathbbm J}$ of coupling constants $J_{ij}$
(with vanishing diagonal entries) and introduce the {\em dressed coupling matrix} ${\mathbbm J} ({\boldsymbol\lambda})$ defined by
\begin{equation}\label{defJdrssed}
 {\mathbbm J} ({\boldsymbol\lambda })_{ij}:=J_{ij}+\lambda_i\,\delta_{ij}\quad\mbox{for}\quad i,j=1,\ldots,N
 \;,
\end{equation}
with  variable diagonal entries $\lambda_i$  subject to the condition $\sum_i \lambda_i=0$.
The choice of ${\boldsymbol\lambda}$ will be often referred to as a {\em gauge}.
The energy of the spin configuration is independent of the gauge.
The minimal eigenvalue of ${\mathbbm J} ({\boldsymbol\lambda})$ will be denoted by $j_{\rm min}({\boldsymbol\lambda})$
and the absolute ground state energy is given by
\begin{equation}\label{gsenergy}
 E_{\rm min}/N =j_{\rm min}(\widehat{\boldsymbol\lambda})
  \;,
\end{equation}
where $\widehat{\boldsymbol\lambda}$ is the {\em ground state gauge}.
Another useful gauge is the {\em homogeneous gauge} $\widetilde{\boldsymbol\lambda}$ defined as follows.
Consider the mean row sum of the undressed  $J$-matrix ${\mathbbm J}$
\begin{equation}\label{meanrs}
 \tilde{\jmath}:=\frac{1}{N}\sum_{ij} J_{ij}
 \;,
\end{equation}
and define
\begin{equation}\label{defhomgauge}
 \widetilde{\lambda}_i:=\tilde{\jmath}-\sum_{j} J_{ij}\quad\mbox{for}\quad i=1\ldots,N
 \;.
\end{equation}
It follows that $ \tilde{\jmath}$ will be an eigenvalue of  ${\mathbbm J} ( \widetilde{\boldsymbol\lambda})$
corresponding to the eigenvector $(1,1,\ldots,1)$ representing a fully aligned spin configuration.

Then the saturation field is given by
\begin{equation}\label{hsata}
 H_{\rm sat}=2\left(\tilde{\jmath}-j_{\rm min}(\widetilde{\boldsymbol\lambda})\right)
 \;,
\end{equation}
see  \cite{HJS_classical_IV}.
Especially, if $\tilde{\jmath}=j_{\rm min}(\widetilde{\boldsymbol\lambda})$ and hence  $H_{\rm sat}=0$ the spin system will assume the ferromagnetic
ground state $\uparrow\uparrow\ldots\uparrow$ even for vanishing fields and hence has been called ``ferromagnetic" in \cite{HJS_classical_IV}.

As an example we consider the AF square-kagom\'{e} with $J_1=J_2=1$ and $J_\times,J_+\ge 0$. It can be shown that
\begin{equation}\label{hsatSK}
H_{\rm sat}=\left\{
\begin{array}{r@{\quad:\quad}l}
2 (3+J_{\times}) & J_{\times}\ge 2\,J_+,\\
 4+J_\times+2 J_++\sqrt{\left(J_\times-2 J_+\right){}^2+4}&J_{\times}\le 2\,J_+,\\
\end{array}
\right.
\;.
\end{equation}
For the special case $J_+=J_\times=:J_3\ge 0$ considered in this paper this equation specializes to
\begin{equation}\label{hsatSK3}
H_{\rm sat}=4 + 3 J_3 + \sqrt{4 + J_3^2}
\;.
\end{equation}

The analogous result for the AF/FM square-kagom\'{e} with $J_1=-J_2=1$ and $J_+,J_\times\ge 0$ reads
\begin{equation}\label{hsatSKF}
H_{\rm sat}=\left\{
\begin{array}{r@{\quad:\quad}l}
2 (1+J_{\times}) & J_{\times}\ge 2\,J_+ -2,\\
{J_\times}+2 {J_+}-2+\sqrt{({J_\times}-2 {J_+}+2)^2+4} & J_{\times}\le2\,J_+ -2,\\
\end{array}
\right.
\;.
\end{equation}
For the special case $0\le J_+=J_\times=:J_3\le 2$ considered in the examples in this paper this equation specializes to
\begin{equation}\label{hsatSKF1}
H_{\rm sat}=2\left(1+J_3\right)
\;.
\end{equation}

\section{Sum rule for the magnetization curve of general Heisenberg spin systems}\label{sec:SRG}

In this section we will formulate and prove some generalization of the results in Section \ref{sec:SR} for general classical
Heisenberg spin systems.
Some of these results already appear in \cite{HJS_classical_IV}. The notation to be introduced only applies to this section.
Analogous results apply to quantum spin systems,
but here the magnetization curve is a step function and its area can be expressed as a sum.
This explains the wording ``sum rule".

We consider the magnetization per site $\mu=M/N$ as a function $\mu(h)$ of the magnetic field $h$ for $0\le h \le h_{\rm sat}$
and its inverse function by $h(\mu)$.
The minimal energy per site (without Zeeman term) for given $\mu$ will be denoted by $\epsilon(\mu):=E_{\rm min}(M)/N$.
It follows that
\begin{equation}\label{hmu}
  h(\mu) = \frac{\partial \epsilon(\mu)}{\partial \mu}
  \;.
\end{equation}

The total energy per site (with Zeeman term) will be denoted by
\begin{equation}\label{Hh}
{\sf H}(h) := \epsilon(\mu(h)) - \mu(h)\,h
\;.
\end{equation}
Hence ${\sf H}(h)$ is the Legendre transform of $\epsilon(\mu)$, see  \cite{HJS_classical_IV},
and
\begin{equation}\label{Hmu}
 \frac{\partial {\sf H}(h) }{\partial h}=-\mu(h)
 \;.
\end{equation}

{\em Cum grano salis}, this also applies if $\mu(h)$ is not differentiable for a finite number of points,
as in the examples of exotic magnetization curves considered in this paper, see Figure \ref{FIGMC}.

The area $F$ below  the graph of $\mu(h)$ is given by
\begin{eqnarray}
\label{F1}
  F &=& \int_{0}^{h_{\rm sat}}\mu(h)\,dh \stackrel{(\ref{Hmu})}{=}- \int_{0}^{h_{\rm sat}}  \frac{\partial {\sf H}(h) }{\partial h}\,dh
  ={\sf H}(0)-{\sf H}(h_{\rm sat}) \\
  \label{F2}
  &=& \epsilon(\mu(0))-\epsilon(\mu(h_{\rm sat})) +\mu(h_{\rm sat})\,h_{\rm sat}=\epsilon(\mu(0))-\epsilon(1)+h_{\rm sat}
  \;,
\end{eqnarray}
since $\mu(h_{\rm sat})=1$.
The complementary area $G$ above the graph of $\mu(h)$, see Figure \ref{FIGMC}, is given by
\begin{equation}\label{G1}
 G:= \int_{\mu(0)}^{1}h(\mu)\,d\mu= h_{\rm sat}-F = -\epsilon(\mu(0))+\epsilon(1)
 \;.
\end{equation}

Recall the definitions and results given in Appendix \ref{sec:SF}.
Then the absolute ground state energy is given by
\begin{equation}\label{gsenergy}
  \epsilon(\mu(0))=j_{\rm min}(\widehat{\boldsymbol\lambda})
  \;,
\end{equation}
where $\widehat{\boldsymbol\lambda}$ is the ground state gauge.
Moreover, for the fully aligned FM state we obtain
\begin{equation}\label{meanrs}
 \epsilon(1)=\frac{1}{N}\sum_{ij} J_{ij} =\tilde{\jmath}
 \;,
\end{equation}
the mean row sum of the undressed $J$-matrix.
Recall that the saturation field is given by
\begin{equation}\label{hsata}
 h_{\rm sat}=2\left(\tilde{\jmath}-j_{\rm min}(\widetilde{\boldsymbol\lambda})\right)
 \;,
\end{equation}
see  (\ref{hsata}). This entails
\begin{eqnarray}
\label{F3}
  F &\stackrel{(\ref{F2})}{=}& \epsilon(\mu(0))-\epsilon(1)+h_{\rm sat}\stackrel{(\ref{gsenergy},\ref{meanrs},\ref{hsata})}{=}
  j_{\rm min}(\widehat{\boldsymbol\lambda})-\tilde{\jmath}+2\left(\tilde{\jmath}-j_{\rm min}(\widetilde{\boldsymbol\lambda})\right)
   \\
\label{F4}
   &=& \frac{1}{2}\,h_{\rm sat}+\left( j_{\rm min}(\widehat{\boldsymbol\lambda})-j_{\rm min}(\widetilde{\boldsymbol\lambda})\right)
   \;.
\end{eqnarray}
In the case $j_{\rm min}(\widehat{\boldsymbol\lambda})=j_{\rm min}(\widetilde{\boldsymbol\lambda})$ it follows that
$F=G= \frac{1}{2}\,h_{\rm sat}$. Especially this holds in the ``parabolic" case where $\mu(h)$ is linear, i.~e.,
 $\mu(h)=\frac{h}{h_{\rm sat}}$ (dotted line in Figure \ref{FIGMC}). In general, we have $F\ge \frac{1}{2}\,h_{\rm sat}$,
 since $j_{\rm min}(\widehat{\boldsymbol\lambda})\ge j_{\rm min}(\widetilde{\boldsymbol\lambda})$, see \cite{HJS_classical_I}.
 The condition $j_{\rm min}(\widehat{\boldsymbol\lambda})> j_{\rm min}(\widetilde{\boldsymbol\lambda})$, or, equivalently,
 $F > \frac{1}{2}\,h_{\rm sat}$ is hence an indication for an exotic magnetization curve, but,
 strictly speaking, it is neither necessary nor sufficient.

\subsection{Sum rule for the AF  square-kagom\'{e}}\label{sec:SRAF}

In order to apply this to the AF  square-kagom\'{e} with $J_1=J_2=1$ and $J_\times,J_+\ge 0$ we have to make a case distinction.

\subsubsection{Case of $J_\times < 2 J_+$}\label{sec:SRAF1}

We have the following results:
\begin{eqnarray}
\label{AF1}
  j_{\rm min}(\widehat{\boldsymbol\lambda})&=&-1- \frac{1}{3} \left(J_\times+J_+\right), \\
  \label{AF2}
  j_{\rm min}(\widetilde{\boldsymbol\lambda}) &=& \frac{1}{6} \left(-3 \sqrt{(2J_+ - J_\times)^2+4}-4 J_+-J_\times\right), \\
  \label{AF3}
  \Delta F &:=& F-\frac{1}{2}\,h_{\rm sat}=
 -1+ \frac{1}{2} \sqrt{(2J_+ - J_\times)^2+4}+\frac{1}{6}\left( 2J_+-J_\times\right) > 0
   \;,
\end{eqnarray}
thus indicating exotic magnetization curves for all $J_\times < 2\,J_+$.
(\ref{AF1}) follows from eq.~(9) in Ref.~\cite{squago_clas_2023} and (\ref{gsenergy}).

\subsubsection{Case of $J_\times \ge 2 J_+$}
We have the following results:
\begin{eqnarray}
\label{AF4}
  j_{\rm min}(\widehat{\boldsymbol\lambda})&=&-1- \frac{1}{3} \left(J_\times+J_+\right), \\
  \label{AF5}
  j_{\rm min}(\widetilde{\boldsymbol\lambda}) &=& -1+\frac{1}{3}\left(J_+-2 J_\times\right), \\
  \label{AF6}
  \Delta F &:=& F-\frac{1}{2}\,h_{\rm sat}=
  \frac{1}{3} \left( J_\times-2\,J_+\right)\ge 0
   \;,
\end{eqnarray}
thus indicating exotic magnetization curves for all $J_\times > 2\,J_+$.

\subsection{Sum rule for the AF/FM  square-kagom\'{e}}\label{sec:SRF}

To state the analogous results for the AF/FM  square-kagom\'{e} with $J_1=-J_2=1$ and $J_\times,J_+\ge 0$ we make the following case distinction.

\subsubsection{Case of $J_\times > 2 J_+-2$}\label{sec:SRAF1}

We have the following results:
\begin{eqnarray}
\label{AF7}
  j_{\rm min}(\widehat{\boldsymbol\lambda})&=&-1- \frac{1}{3} \left(J_\times+J_+\right), \\
  \label{AF8}
  j_{\rm min}(\widetilde{\boldsymbol\lambda}) &=&  \frac{1}{3} \left(-5 -2J_\times+J_+\right), \\
  \label{AF9}
  \Delta F &:=& F-\frac{1}{2}\,h_{\rm sat}=
\frac{1}{3} \left(J_\times-2\,J_+ +2\right) > 0
   \;,
\end{eqnarray}
thus indicating exotic magnetization curves for all $J_\times > 2\,J_+-2$.
(\ref{AF7}) is identical to (\ref{AF1}) due to the fact that the AF and the AF/FM  square-kagom\'{e}
have the same ground state energy (without Zeeman term).

\subsubsection{Case of $J_\times \le 2 J_+-2$}\label{sec:SRAF2}
We have the following results:
\begin{eqnarray}
\label{AF10}
  j_{\rm min}(\widehat{\boldsymbol\lambda})&=&-1- \frac{1}{3} \left(J_\times+J_+\right), \\
  \label{AF11}
  j_{\rm min}(\widetilde{\boldsymbol\lambda}) &=&\frac{1}{6} \left(-3 \sqrt{(-{J_\times}+2 {J_+}-2)^2+4}-J_\times-4 {J_+}+2\right), \\
  \label{AF12}
  \Delta F &:=& F-\frac{1}{2}\,h_{\rm sat}=
  \frac{1}{6} \left(3 \sqrt{(J_\times-2 {J_+}+2)^2+4}-J_\times+2 {J_+}-8\right)\ge 0
   \;,
\end{eqnarray}
thus indicating exotic magnetization curves for all $J_\times < 2\,J_+-2$.

\section{AF square-kagom\'{e}  phase IV }\label{sec:AIV}

It is possible to describe phase IV analytically for all values of $J_3>0$ if we assume that the ground state has the fxorm described in the main text
depending on two $z$-values $z_1$ and $z_2$.
Its energy $E(J_3,H,z_1,z_2)$ is given by
\begin{equation}\label{enhz12}
 E(J_3,H,z_1,z_2)=\frac{1}{3} \left(-H \left(z_1+2 z_2\right)+2 {J_3}
   \left(z_1^2+z_2^2-1\right)+2 z_2^2+4 z_1 z_2-2
   \sqrt{2-2 z_1^2} \sqrt{1-z_2^2}\right)
   \;.
\end{equation}
Ground states of this phase satisfy  $\partial E/\partial z_1=\partial E/\partial z_2=0 $.
These equations admit a parametric solution of the following form:
\begin{eqnarray}
\label{parsol1da}
  z_1&=&\frac{w \left(\sqrt{2} \left(J_3-1\right)+w\right)}{\sqrt{2} \left(2 J_3-1\right) w+2}\,W\\
  \label{parsol1db}
  z_2 &=& W\\
  M &=&\frac{\sqrt{2} \left(5 J_3-3\right) w+w^2+4}{3 \sqrt{2} \left(2 J_3-1\right) w+6}\,W\\
  \label{parsol1dc}
  H &=&\frac{2 \left(\sqrt{2} J_3 w^2+\left(2 J_3 \left(J_3+1\right)-1\right) w+\sqrt{2}
   \left(J_3+1\right)\right)}{\left(2 J_3-1\right) w+\sqrt{2}}\,W, \mbox{ where }\\
  \label{parsol1dd}
  W&:=&\frac{\sqrt{\left(w^2-1\right) \left(\left(2 J_3-1\right) w \left(\left(2 J_3-1\right) w+2
   \sqrt{2}\right)+2\right)}}{w \sqrt{\frac{1}{2} \left(8 \left(J_3-1\right) J_3+1\right)
   w^2+\sqrt{2} \left(3 J_3-1\right) w-\left(J_3-2\right) J_3+1}}
  \;,
\end{eqnarray}
and $w$ is a parameter satisfying $1\le w \le w_{max}(J_3)$.
For example, $w_{max}(J_3=1)=\frac{1+\sqrt{5}}{\sqrt 2}=2.28825\ldots$.


\end{document}